\documentclass[aps,prd,twocolumn,superscriptaddress,showpacs,nofootinbib,preprintnumbers]{revtex4-1}

\usepackage{amsmath,amssymb,graphicx}
\usepackage{color}
\usepackage{pslatex}
\usepackage{graphicx}
\usepackage{psfrag}

\usepackage{graphicx}
\usepackage{hyperref}

\usepackage[normalem]{ulem}

\begin{document}

\title{Unique probe of dark matter in the core of M87 with the Event Horizon Telescope}

\author{Thomas Lacroix}
\affiliation{Laboratoire Univers \& Particules de Montpellier (LUPM), CNRS \& Universit\'{e} de Montpellier (UMR 5299), Place Eug\`{e}ne Bataillon, F-34095 Montpellier Cedex 05, France}
\affiliation{Institut d'Astrophysique de Paris, UMR 7095, CNRS, UPMC Universit\'{e} Paris 6, Sorbonne Universit\'{e}s, 98 bis boulevard Arago, 75014 Paris, France}
\email{thomas.lacroix@umontpellier.fr}
\author{Mansour Karami}
\affiliation{Perimeter Institute for Theoretical Physics, 31 Caroline Street North, Waterloo, Ontario N2L 2Y5, Canada}
\affiliation{Department of Physics and Astronomy, University of Waterloo, 200 University Avenue West, Waterloo, ON N2L 3G1, Canada}
\author{Avery E. Broderick}
\affiliation{Perimeter Institute for Theoretical Physics, 31 Caroline Street North, Waterloo, Ontario N2L 2Y5, Canada}
\affiliation{Department of Physics and Astronomy, University of Waterloo, 200 University Avenue West, Waterloo, ON N2L 3G1, Canada}
\author{Joseph Silk}
\affiliation{Institut d'Astrophysique de Paris, UMR 7095, CNRS, UPMC Universit\'{e} Paris 6, Sorbonne Universit\'{e}s, 98 bis boulevard Arago, 75014 Paris, France}
\affiliation{The Johns Hopkins University, Department of Physics and Astronomy,
3400 N. Charles Street, Baltimore, Maryland 21218, USA}
\affiliation{Beecroft Institute of Particle Astrophysics and Cosmology, Department of Physics,
University of Oxford, Denys Wilkinson Building, 1 Keble Road, Oxford OX1 3RH, United Kingdom}
\author{C\'{e}line B{\oe}hm}
\affiliation{Institute for Particle Physics Phenomenology, Durham University, Durham, DH1 3LE, United Kingdom}
\affiliation{LAPTH, Universit\'{e} de Savoie, CNRS, BP 110, 74941 Annecy-Le-Vieux, France}

\date{\today}

\begin{abstract}
We demonstrate the unprecedented capabilities of the Event Horizon Telescope (EHT) to image the innermost dark matter profile in the vicinity of the supermassive black hole at the center of the M87 radio galaxy. We present the first model of the synchrotron emission induced by dark matter annihilations from a spiky profile in the close vicinity of a supermassive black hole, accounting for strong gravitational lensing effects. Our results show that the EHT should readily resolve dark matter spikes if present. Moreover, the photon ring surrounding the silhouette of the black hole is clearly visible in the spike emission, which introduces observable small-scale structure into the signal. We find that the dark matter-induced emission provides an adequate fit to the existing EHT data, implying that in addition to the jet, a dark matter spike may account for a sizable portion of the millimeter emission from the innermost (subparsec) region of M87. Regardless, our results show that the EHT can probe very weakly annihilating dark matter. Current EHT observations already constrain very small cross sections, typically down to a few $10^{-31}\ \mathrm{cm^{3}\ s^{-1}}$ for a 10 GeV candidate, close to characteristic values for p-wave-suppressed annihilation. Future EHT observations will further improve constraints on the DM scenario.
\end{abstract}

\pacs{95.35.+d, 96.50.S-, 95.55.Br}

\preprint{LUPM:16-020}

\maketitle

\section{Introduction}

The dark matter (DM) density profile at the centers of galaxies is critical to indirect searches but remains poorly constrained. In objects such as M87, the DM profile may be significantly enhanced on subparsec scales by the central supermassive black hole (SMBH), although there is no direct evidence for such a sharp density increase, referred to as a DM spike.

DM spikes however leave a very distinctive signature in the spectral energy distribution (SED) of a galaxy if DM can annihilate \cite{spikeGS,M87_limits_my_paper,Fields2014}. More specifically, in the case of M87, where a spike should plausibly have formed and survived galaxy dynamics, an anomalous contribution to the SED is expected when the DM annihilation cross section is larger than $\sim 10^{-30}\ \rm cm^{3}\ s^{-1}$ for light (10 to 100 GeV) DM particles, and above $\sim 3 \times 10^{-26}\ \rm cm^{3}\ s^{-1}$ for candidates as heavy as 100 TeV \cite{M87_limits_my_paper}.

Here we show that it is possible to probe even fainter DM-induced radiation in M87 by using the spatial morphology of the DM-induced synchrotron emission near the central black hole (BH). Due to a lack of angular resolution in existing observational facilities, such a study of the DM-induced signal in the inner part of M87 has not been performed yet---nor has it been done in similar objects. However, this is now possible with the advent of the Event Horizon Telescope (\href{http://www.eventhorizontelescope.org/}{EHT}).

Heavy DM particles annihilating into Standard Model particles near the central BH are expected to produce synchrotron emission in the frequency range that is currently being probed by the EHT. Here, we show that the synchrotron halo induced by DM is bright enough to be resolved by the EHT, if there is a DM spike. Moreover, this additional radiation enhances the photon ring around the BH shadow, thus making it a prominent feature in the EHT data and a new probe of the DM properties.

In Sec.~\ref{EHT}, we provide a description of the EHT, and then we present our DM model in Sec.~\ref{DM&EHT} and the EHT data we used in Sec.~\ref{data}. Our results can be found in Sec.~\ref{results}, and we conclude in Sec.~\ref{conclusion}.

\section{The Event Horizon Telescope}
\label{EHT}

\subsection{General features}

The EHT is a global network of millimeter and submillimeter facilities that employs very long baseline interferometry (VLBI) to create an effective Earth-scale high angular resolution telescope \citep{Doeleman2010,EHT_M87}. The purpose of this array is to test general relativity and shed light on physical processes taking place in the vicinity of SMBHs at the centers of galaxies. To date, EHT data for M87 has been reported for a three-station array comprised of the Submillimeter Telescope (\href{http://aro.as.arizona.edu/smt_docs/smt_telescope_specs.htm}{SMT}) in Arizona, the Combined Array for Research in Millimeter-wave Astronomy (\href{https://www.mmarray.org/}{CARMA}) in California, and a network of three facilities in Hawaii: the James Clerk Maxwell Telescope (\href{http://www.eaobservatory.org/jcmt/}{JCMT}), the Submillimeter Array (\href{https://www.cfa.harvard.edu/sma/}{SMA}), and the Caltech Submillimeter Observatory (\href{http://cso.caltech.edu/}{CSO}). This configuration has already achieved an impressive angular resolution of order $40\ \mu \rm as$ at 230 GHz. Presently, the Atacama Large Millimeter/submillimeter Array (\href{http://www.almaobservatory.org/}{ALMA}) in Chile, the Large Millimeter Telescope (\href{http://www.lmtgtm.org/}{LMT}) in Mexico, the \href{http://www.iram-institute.org/EN/30-meter-telescope.php}{Institut de Radioastronomie Millim\'{e}trique (IRAM) 30m}, the \href{http://www.iram-institute.org/EN/plateau-de-bure.php?ContentID=3&rub=3&srub=0&ssrub=0&sssrub=0}{Plateau de Bure interferometer}, and the South Pole Telescope (\href{http://pole.uchicago.edu/}{SPT})\footnote{Note that M87 cannot be seen by the SPT.} will be added to the EHT.  Longer term, the \href{https://www.cfa.harvard.edu/greenland12m/telescope/}{Greenland Telescope} will join the array. Combined, the EHT will directly access angular scales as small as $26\ \mu \rm as$ at 230 GHz and $17\ \mu \rm as$ at 345 GHz.  

The angular scale of the Schwarzschild radius of the SMBH at the center of M87 is about $8\ \mu \mathrm{as}$.\footnote{We use $M_{\mathrm{BH}} = (6.4 \pm 0.5) \times 10^9 M_{\odot}$ \cite{Gebhardt2009} for the mass of the central BH. This value is based on stellar dynamics measurements, and is consistent with other more recent similar estimates \cite{Gebhardt2011,Oldham2016}. The corresponding Schwarzschild radius is $R_{\mathrm{S}} = 6 \times 10^{-4}\ \rm pc$ and the distance of M87 is $d_{\mathrm{M87}} \approx 16\ \mathrm{Mpc}$ \citep{Bird2010}. Throughout we adopt the higher stellar dynamical mass for M87; using the $(3.5\pm0.8)\times10^9 M_\odot$ value found by gas dynamical studies \cite{Walsh2013} would reduce the angular scales by roughly a factor of 2 througout. This would make it more difficult to detect horizon-scale features with the EHT. However, we believe stellar dynamics measurements to be more reliable than gas dynamical studies since the latter make rather extreme assumptions on the kinematical properties of the gas---such as considering that it moves on circular, Keplerian orbits---, or using a simplified disk model for the gas, unlikely to account for the high velocity dispersions derived from line measurements.} The strong gravitational lensing near the black hole (BH) magnifies this by a factor of as much as 2.5, only weakly dependent on BH spin, making M87 a primary target for the EHT \citep{EHT_M87}.

\subsection{Imaging the shadow of a black hole}

Characteristic in all images of optically thin emission surrounding BHs is a ``shadow''---a dark central region surrounded by a brightened ring, the so-called ``photon ring''---. This is a direct consequence of the strong gravitational lensing near the photon orbit, and is directly related to the projected image of the photon orbit at infinity. The shadow interior is the locus of null geodesics that intersect the horizon, and thus does not contain emission from behind the BH. The bright ring is in general sharply defined due to the instability of the photon orbit and a consequence of the pileup of higher-order images of the surrounding emission. For a Schwarzschild BH the radius of this shadow is $r_{\mathrm{shadow}} = 3\sqrt{3}/2 R_{\mathrm{S}} \approx 2.6 R_{\mathrm{S}}$ \cite{Bardeen1973,Bozza2010}. For Kerr BHs, this radius ranges from $2.25 R_{\mathrm{S}}$ to $2.6 R_{\mathrm{S}}$, deviating substantially only at large values of the dimensionless spin parameter and viewed from near the equatorial plane, i.e., $a\gtrsim0.9$ and $\theta\gtrsim60^\circ$ \cite{Johannsen2011}.

The generic appearance of the shadow, its weak dependence on BH spin, and fundamentally general-relativistic origin make it a prime feature in EHT science. Also imprinted on EHT images will be the high-energy astrophysics of the near-horizon region; the physics of BH accretion and relativistic jet formation. Horizon-scale features have already been observed in early EHT observations of Sagittarius A* (Sgr A*) \cite{Doeleman2008,Fish2011,Fish2016} and M87 \citep{Doeleman2012,Akiyama2015}, demonstrating that such structure exists. Here, we assess the observability of the shadow of the SMBH at the center of M87 in the electromagnetic signal from DM annihilation, and the limits that may be placed on DM properties given such a scenario.

\section{Probing a dark matter spike at the center of M87 with the Event Horizon Telescope}
\label{DM&EHT}

\subsection{Ingredients and assumptions}

In the context of the observational opportunities offered by the EHT, we now discuss the potential of this instrument in terms of DM searches. In particular, we study the observability of a DM spike at the center of M87, since such a sharply peaked morphological feature is expected to yield strong annihilation signals. At the frequencies of interest for the EHT, typically a few hundred GHz, the main DM signature comes from synchrotron radiation. Therefore, in order to assess the ability of the EHT to probe the inner part of the DM profile of M87, we need to compute the synchrotron emission of electrons and positrons produced in DM annihilations in the inner region. We make the following assumptions:

\begin{itemize}
\item The presence of a SMBH at the center is likely to lead to fairly strong magnetic fields, typically around 10--$10^{2}$ G \cite{Broderick2015}. As a result of such strong magnetic fields, synchrotron radiation and advection towards the central BH are the dominant physical processes by which DM-induced electrons and positrons lose or gain energy \citep{Aloisio2004,Regis&Ullio}, whereas inverse Compton scattering and bremsstrahlung are negligible. Additionally, the time scales associated with synchrotron radiation and advection are much shorter than that of spatial diffusion, so we disregard the latter in the following. Note that even larger magnetic fields---up to $\sim 10^{3}$ G---would arise if the equipartition scenario proposed in Ref.~\cite{Melia1992} for the center of the Milky Way were realized in M87. Therefore, to account for the uncertainty on the central magnetic field strength, we consider values in the range $10$--$10^{3}\ \mathrm{G}$.

\item We also disregard two processes that can reduce the synchrotron intensity. On the one hand, the synchrotron self-Compton effect, which would lead to additional energy losses for electrons and positrons, is only relevant for magnetic fields significantly smaller than 0.1 G \citep{Aloisio2004}. For larger magnetic fields, such as the ones we consider here, synchrotron self-Compton losses are negligible with respect to synchrotron losses. On the other hand, synchrotron self-absorption is only relevant below $\sim$ 10 GHz \citep{Aloisio2004,Regis&Ullio}, so it can be neglected for the EHT frequency of 230 GHz. 

\item We want to test the presence of a spike in the DM profile, formed through adiabatic growth of a SMBH \cite{spikeGS} at the center of a DM halo with power-law density profile. The existence of such a strong enhancement of the DM density---corresponding to $\rho(r) \propto r^{-\gamma_{\mathrm{sp}}}$ with typically $\gamma_{\mathrm{sp}} = 7/3$---is debated. An adiabatic spike can actually be weakened by various dynamical processes such as mergers \cite{Merritt2002}. It turns out that M87 may contain a BH binary in the central region, considering current evidence for a $\sim 10\ \rm pc$ displacement between the SMBH and the center of the galaxy \cite{Batcheldor2010,Lena2014}, and the discovery of a hypervelocity cluster \cite{Caldwell2014}. This is suggestive of binary scouring at $\sim 10\ \rm pc$ scales. However, here we are actually interested in the DM spike much closer in, which would be unaffected. A softer cusp is also formed if the BH did not grow exactly at the center of the DM halo (within $\sim 50\ \rm pc$) \cite{Nakano1999,Ullio2001}, or if the BH growth cannot be considered adiabatic \cite{Ullio2001}. Moreover, dynamical heating in the central stellar core would also soften a spike \cite{Gnedin2004}. However, unlike in the Milky Way, an adiabatic spike is more likely to have survived in a dynamically young galaxy such as M87, in which stellar heating is essentially negligible---essentially due to the large velocity dispersion caused by the very massive SMBH---, as discussed in Refs.~\cite{Vasiliev2008,M87_limits_my_paper}. Furthermore, other dynamical processes can have the opposite effect of making the survival of a spike more likely, such as enhanced accretion of DM to counteract the depopulation of chaotic orbits in triaxial halos \cite{Merritt2004triaxial}. All these arguments motivate the assumption that a steep spike effectively formed at early times at the center of M87 and has survived until today.\footnote{An additional caveat is related to a putative stellar spike which might have formed jointly with a DM spike in the adiabatic formation scenario. However, this would strongly rely on the existence of a nuclear star cluster (NSC), which in the most accepted view is formed by merging globular clusters \cite{Antonini2015}. Unlike the Milky Way, M87 is actually not an optimal candidate for such mergers, due to the large velocity dispersion induced by the very massive central BH. Moreover, no NSC has been observed in M87 (e.g.~Ref.~\cite{Byun1996}). Therefore, stars and DM essentially decouple in this regard, and the absence of a stellar spike in observations does not preclude the existence of a DM spike.}
\footnote{We note that there is significant uncertainty on the halo profile, which can in principle affect the central density in the spike and the resulting synchrotron fluxes. Here for definiteness we assume that the halo follows the NFW profile, although this is debatable. In Ref.~\cite{Murphy2011} the authors found the data to be consistent with the NFW profile, while the results of Ref.~\cite{Oldham2016} favor a cored generalized NFW-like DM distribution. The authors of Ref.~\cite{Gammaldi2016} discuss the impact of the outer halo on the spike model---especially on the outer radius of the spike $R_{\mathrm{sp}}$---, using the prescription given in Ref.~\cite{spikeGS} (see Appendix \ref{spike_normalization_M87}). They argue that a cored halo would lead to a lower central density due to a smaller value for $R_{\mathrm{sp}}$. However, the prediction for $R_{\mathrm{sp}}$ should not be taken at face value, especially when comparing different halo profiles. It should only be interpreted as a benchmark, all the more so as it can be significantly affected by the various dynamical processes described above. The extent of the spike should actually be of the order of the radius of gravitational influence of the SMBH---of order 100 pc for M87 from the $M_{\mathrm{BH}}-\sigma$ relation \cite{Ferrarese2005}---, regardless of the halo profile. As a result, our conclusions would only be mildly affected by a different choice in the outer halo, provided the spike roughly spans the sphere of influence of the BH and the inner slope is $\gamma_{\mathrm{sp}} \gtrsim 2$.}

\end{itemize}

\subsection{Electron propagation in the presence of advection}

To derive the intensity of DM-induced synchrotron radiation, we first need to compute the electron and positron spectra from the DM annihilation rate. This is done by solving the propagation equation of DM-induced electrons and positrons which, in the presence of synchrotron radiation and advection, and assuming a steady state reads (see e.g.~Refs.~\cite{Aloisio2004,Regis&Ullio})
\begin{equation}
\label{propagation_equation_advection}
v\dfrac{\partial f_{i}}{\partial r} - \dfrac{1}{3 r^{2}} \dfrac{\partial}{\partial r}\left( r^{2} v \right) p \dfrac{\partial f_{i}}{\partial p} + \dfrac{1}{p^{2}} \dfrac{\partial}{\partial p} \left( p^{2} \dot{p} f_{i} \right) = Q_{i},
\end{equation}
where $f_{i}(r,p)$ is the distribution function of electrons and positrons in momentum space, at radius $r$ and momentum $p$, for annihilation channel $i$. The first, second and third terms correspond to the advection current, the energy gain of electrons due to the adiabatic compression, and the loss term due to radiative losses, respectively. $v(r) = - c \left( r/R_{\mathrm{S}} \right)^{-1/2}$ is the radial infall velocity of electrons and positrons onto the BH in the accretion flow, where $R_{\mathrm{S}}$ is the Schwarzschild radius. The minus sign in the expression of the inflow velocity accounts for the direction of the flow, oriented towards the BH.

The source function $Q_{i}(r,p)$ is the DM annihilation rate in momentum space for channel $i$, related to the annihilation rate $q_{i}(r,E)$ in energy space via
\begin{equation}
Q_{i}(r,p) = \dfrac{c}{4 \pi p^{2}} q_{i}(r,E)
\end{equation}
in the ultrarelativistic (UR) regime where $E = p c$. The UR approximation can be safely used for electrons and positrons for the energy range relevant for this study. The usual annihilation rate in energy space reads
\begin{equation}
q_{i}(r,E) = \dfrac{\left\langle \sigma v \right\rangle_{i}}{\eta} \left( \dfrac{\rho(r)}{m_{\mathrm{DM}}} \right) ^{2} \dfrac{\mathrm{d}N_{\mathrm{e},i}}{\mathrm{d}E}(E),
\end{equation}
where $\eta = 2$ for the case of self-annihilating DM that we consider here. The injection spectrum $\mathrm{d}N_{\mathrm{e},i}/\mathrm{d}E$ is taken from Ref.~\cite{Cirelli_cookbook} and the associated \href{http://www.marcocirelli.net/PPPC4DMID.html}{website}.

Since radiative losses are dominated by synchrotron losses, the total radiative loss term $\dot{p} = \mathrm{d}p/\mathrm{d}t$ reduces to the synchrotron loss term \cite{Longair2011} 
\begin{equation}
\label{syn_losses}
\dot{p}(r,p) = \dot{p}_{\mathrm{syn}}(r,p) = - \dfrac{2 \sigma_{\mathrm{T}} B^{2} p^{2}}{3 \mu_{0} (m_{\mathrm{e}} c)^{2}}.
\end{equation}
We assume the intensity of the magnetic field to be homogeneous, i.e.~$B \equiv B_{0}$, over the accretion region, which has a size $r_{\mathrm{acc}}$ corresponding to the sphere of influence of the BH \citep{Regis&Ullio}, so typically $\sim 60\ \rm pc$ (as discussed in Ref.~\cite{M87_limits_my_paper}), which is also roughly the size of the spike.

The resolution of the propagation equation, Eq.~\eqref{propagation_equation_advection}, in the presence of synchrotron losses and advection, in the UR regime, and with the method of characteristics, yields the electron and positron spectrum in terms of the DM annihilation rate \citep{Aloisio2004}
\begin{equation}
f_{i}(r,p) = \dfrac{1}{c} \left( \dfrac{r}{R_{\mathrm{S}}} \right) \int_{r}^{r_{\mathrm{acc}}} \! Q_{i}(R_{\mathrm{inj}},p_{\mathrm{inj}}) \left( \dfrac{R_{\mathrm{inj}}}{R_{\mathrm{S}}} \right)^{\frac{5}{2}} \left( \dfrac{p_{\mathrm{inj}}}{p} \right)^{4} \, \mathrm{d}R_{\mathrm{inj}},
\end{equation}
where the injection momentum $p_{\mathrm{inj}} \equiv p_{\mathrm{inj}}(R_{\mathrm{inj}};r,p) $ for a homogeneous magnetic field is given in Appendix \ref{advection_resolution}. From there, the electron and positron energy spectrum is given by
\begin{equation}
\psi_{i}(r,E) = \dfrac{4 \pi p^{2}}{c} f_{i}(r,p).
\end{equation}
We then convolve $\psi_{i}$ with the synchrotron power $P_{\mathrm{syn}}(\nu,E,r)$ to obtain the synchrotron emissivity:
\begin{equation}
j_{\mathrm{syn},i}(\nu,r) = 2 \int_{m_{\mathrm{e}}}^{m_{\mathrm{DM}}} \ P_{\mathrm{syn}}(\nu,E,r) \psi _{i}(r,E) \, \mathrm{d}E. 
\label{jnu}
\end{equation}

\subsection{Relevance of advection}

Advection shapes the inner part of the intensity profile by displacing electrons and positrons towards the BH, thereby accelerating them. This effect is in competition with synchrotron losses. Therefore, depending on the magnetic field, electrons either lose their energy in place through synchrotron radiation, or are first advected towards the center. The dependence on the magnetic field of the size of the region where electrons are affected by advection is obtained by comparing the synchrotron loss term given in Eq.~\eqref{syn_losses} with the momentum gain rate due to adiabatic compression,
\begin{equation}
\dot{p}_{\mathrm{ad}} = - \dfrac{1}{3 r^{2}} \dfrac{\partial}{\partial r} \left( r^{2} v(r) \right) p.
\end{equation}
The shape of the emissivity profile is thus governed by advection for
\begin{equation}
r \lesssim \left( \dfrac{3 R_{\mathrm{S}}^{\frac{1}{2}} \mu_{0} m_{\mathrm{e}}^{2} c^{4}}{4 \sigma_{\mathrm{T}} B_{0}^{2} E_{\mathrm{syn}}} \right) ^{\frac{2}{3}},
\end{equation}
where $E_{\mathrm{syn}} = \left( 4 \pi m_{\mathrm{e}}^{3} c^{4} \nu/(3 e B_{0}) \right) ^{1/2}$ is the peak synchrotron energy at the frequency $\nu$ of interest \citep{Longair2011}. In terms of the angular distance from the center $\theta \equiv r/d_{\mathrm{M87}}$---where $d_{\mathrm{M87}} \approx 16\ \mathrm{Mpc}$ is the distance between Earth and the center of M87---this condition reads, for $\nu = 230\ \rm GHz$,
\begin{equation}
\theta \lesssim 7 \left( \dfrac{B_{0}}{10\ \mathrm{G}} \right) ^{-1}\ \mu \rm as.
\end{equation}
It is therefore essential to include the advection process for $B_{0} \lesssim 10\ \rm G$, since in that regime it has a strong impact on the synchrotron intensity in the region of interest for the EHT. For $B_{0} \gg 10\ \rm G$, advection is negligible since it would only dominate for radii smaller than the Schwarzschild radius of the BH.

\subsection{Dark synchrotron intensity in a curved spacetime}

The DM-induced synchrotron intensity for a flat spacetime is computed by integrating the emissivity over the line of sight. However, the actual spacetime accounting for the presence of the SMBH at the center is characterized by the Schwarzschild or Kerr metric respectively if the BH is static or rotating. In these realistic cases, the correct spatial morphology of the synchrotron intensity $I_{\nu}$ is obtained using a ray-tracing technique that accounts for the gravitational lensing effect due to the BH. In our case, this is achieved via the ray-tracing and radiative transfer scheme described in Refs.~\cite{Broderick2006,Broderick2006a,Broderick2006b}.

\begin{figure*}[t!]
\centering
\includegraphics[width=0.49\linewidth]{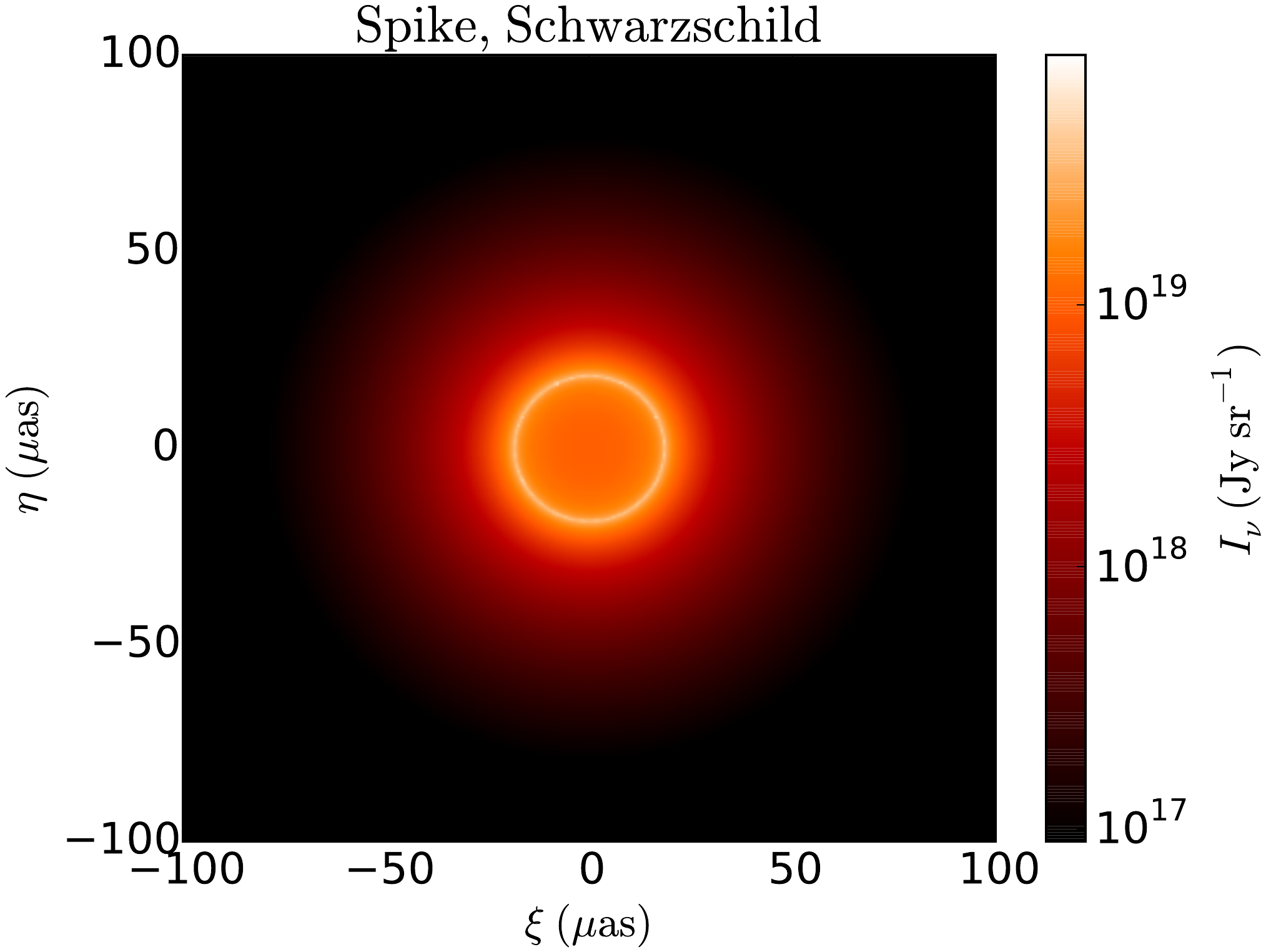} \hfill \includegraphics[width=0.49\linewidth]{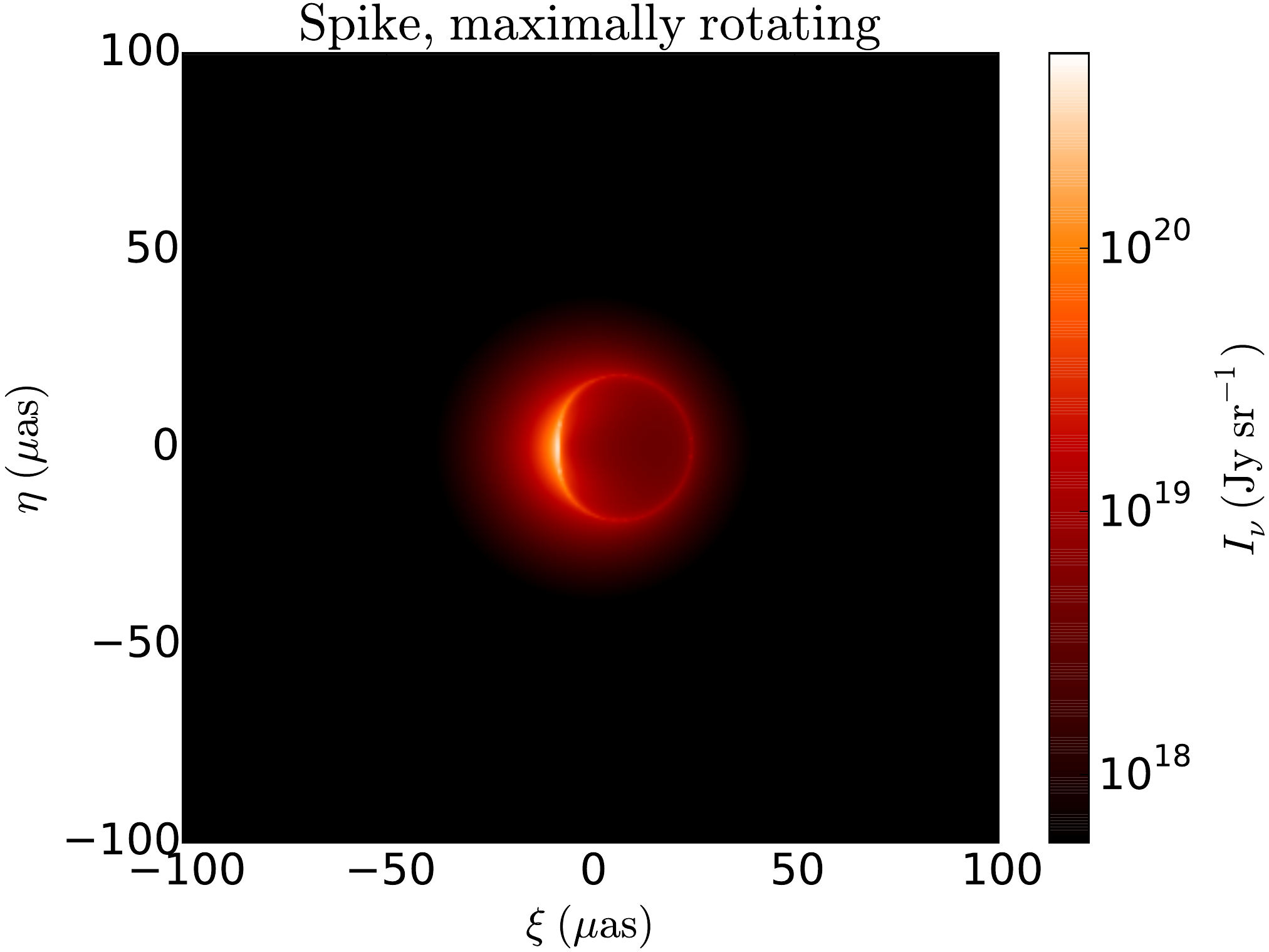} 
\includegraphics[width=0.49\linewidth]{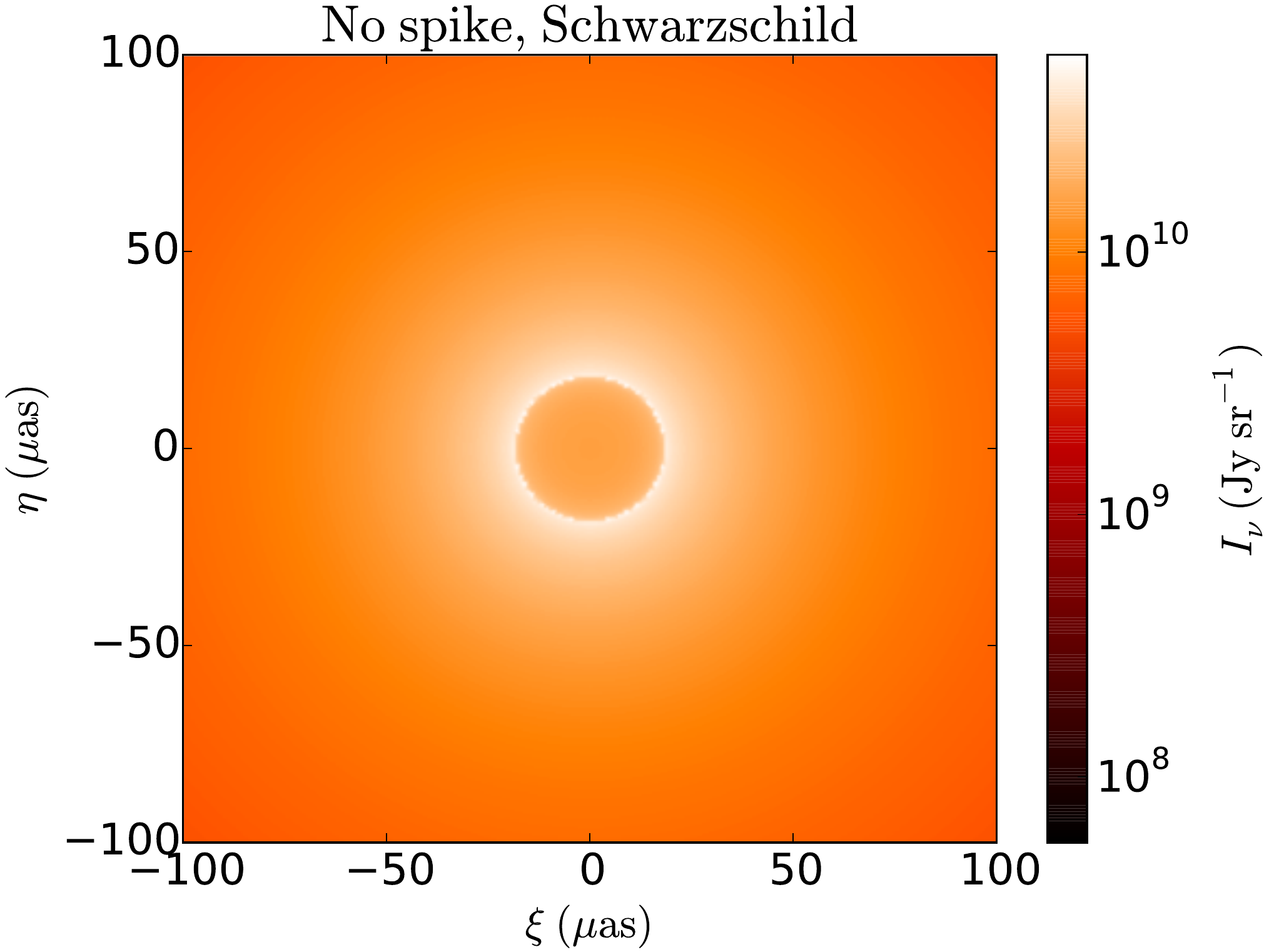}
\hfill \includegraphics[width=0.49\linewidth]{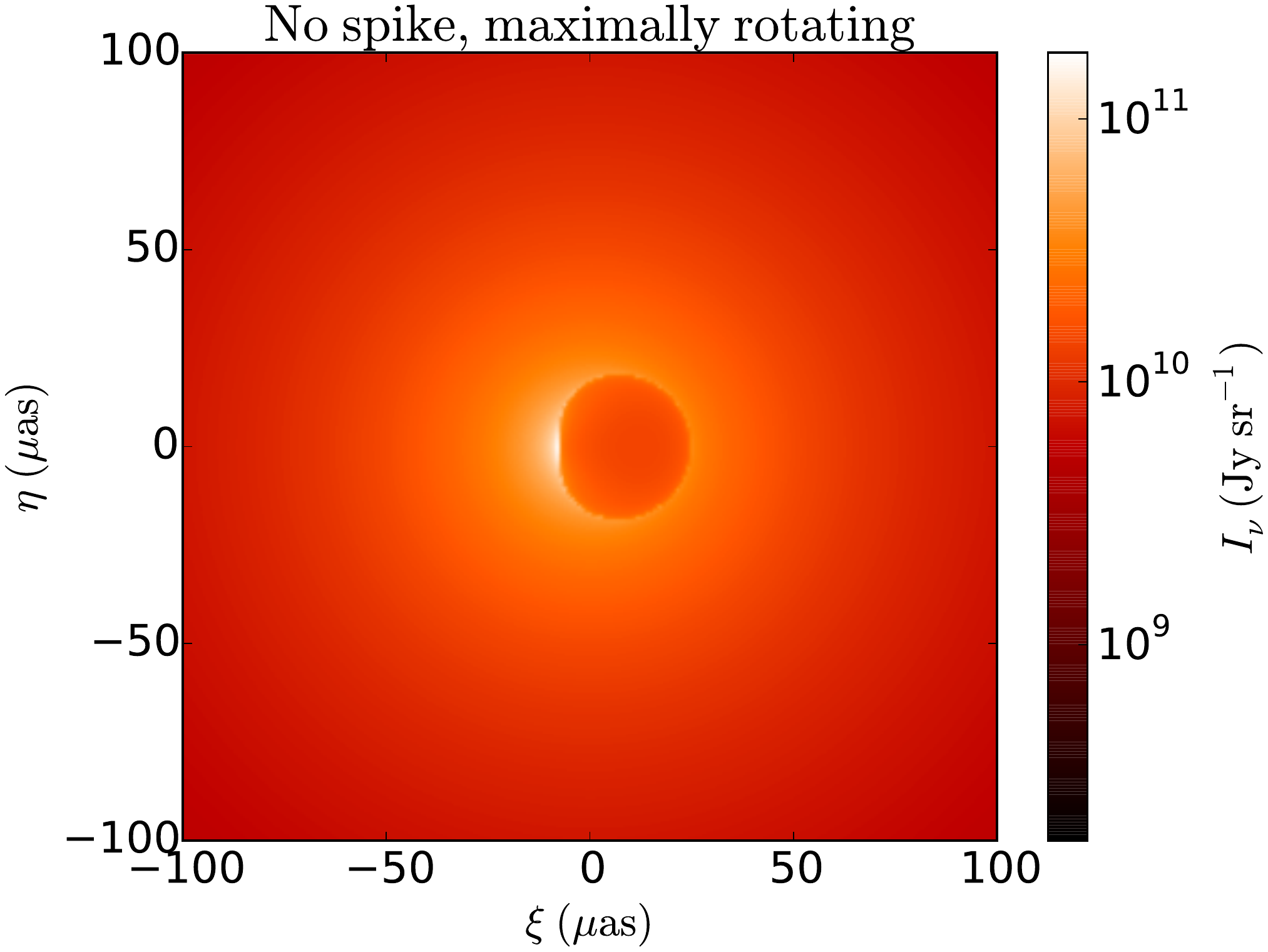}
\caption{\label{shadow_maps}Simulated maps of the synchrotron intensity at 230 GHz from a spike of 10 GeV DM annihilating into $b\bar{b}$, accounting for the strong gravitational lensing induced by the central BH, for a Schwarzschild BH (left panel) and a maximally rotating BH (right panel), in the presence (upper panels) and absence (lower panels) of a spike in the DM profile. Note that considering the wide range of intensities, we use different color scales, but with the same dynamic range spanning 3 orders of magnitude for comparison. The angular coordinates $\xi$ and $\eta$ correspond to the directions respectively perpendicular and parallel to the spin of the BH. For the spike cases, the slope of the DM spike is $\gamma_{\mathrm{sp}} = 7/3$, and the annihilation cross sections correspond to the best fit to EHT observations (see text for details), namely $7.4 \times 10^{-31}\ \mathrm{cm^{3}\ s^{-1}}$ for the Schwarzschild case and $3.1 \times 10^{-31}\ \mathrm{cm^{3}\ s^{-1}}$ for the maximally rotating case. In the absence of a spike, the intensity is computed for the thermal s-wave cross section of $3 \times 10^{-26}\ \mathrm{cm^{3}\ s^{-1}}$. For all the simulated maps the magnetic field is 10 G.}
\end{figure*}

\section{Data}
\label{data}

Here we provide a brief overview of the EHT data used to constrain our models. Details regarding individual observing runs, data calibration, and the uncertainty estimates can be found in Refs.~\cite{Doeleman2012,Akiyama2015}, to which we direct interested readers.

The EHT, like all interferometers, directly constructs visibilities by cross-correlating observations at pairs of stations.  These are directly proportional to the Fourier transform of the image at a spatial frequency proportional to the ratio of the projected baseline distance between the two stations to the observation wavelength.  Throughout the night the rotation of the Earth results in a rotation of the projected baseline, changing both its orientation and length, thereby generating a moderate variation in the spatial frequencies probed by any particular pair of sites.

Importantly, like any interferometer the EHT acts as a ``high-pass'' filter, sensitive primarily to structures on angular scales that lie between $\lambda/u_{\rm short}$ and $\lambda/u_{\rm long}$, where $u_{\rm short}$ and $u_{\rm long}$ are the shortest and longest baselines in the array, respectively.  Within the context of the published EHT data on M87 these angular scales are $75$ to $450$~$\mu$as, though the high signal-to-noise ratio of the data extends these by roughly a factor of 2. Therefore, features that extend over more than a milliarcsecond are effectively invisible to the EHT.

Upon measuring a sufficient number of these visibilities an image can be produced via inverting the Fourier transform.  In practice, this is performed via a number of sophisticated image-inversion techniques that impose additional requirements on the final image, e.g., positivity, smoothness, etc \cite{Lu2014}.  However, because the published EHT observations only sparsely sample the spatial-frequency plane (often called the ``u-v'' plane) we compare directly with the measured visibilities.

By construction the visibilities are complex valued, and therefore described by an amplitude and phase.  However, in practice the amplitudes are known much better than the phases as a result of the typically large, and highly variable, atmospheric phase delays.  This does not mean that phase information is completely unavailable; ``closure phases'' constructed from triplets of sites, equal to the sum of the phases over the closed triangle of baselines, are insensitive to site-specific phase errors.  Therefore, the two data sets we employ consist of visibility amplitudes and closure phases obtained with the Hawaii-SMT-CARMA array, reported in Refs.~\cite{Doeleman2012,Akiyama2015}. In all cases observations were taken at 230~GHz.

Altogether, 160 visibility amplitudes were constructed from data taken on April 5, 6, and 7, 2009, reported in Ref.~\cite{Doeleman2012}, and March 21, 2012, reported in Ref.~\cite{Akiyama2015}.  All nights showed visibility amplitudes consistent with a single source structure.  Of these 54 visibilities were reported on the CARMA-SMT baseline, 83 on the Hawaii-CARMA baseline, and 23 on the Hawaii-SMT baseline.  Note that in all of the reported measurements a 5\% systematic calibration error has been added to the uncertainties in quadrature.

On the Hawaii-CARMA-SMT triangle, 17 closure phases were constructed from data taken on March 21, 2012 and reported in Ref.~\cite{Akiyama2015}.  Where visibility amplitudes provide a measure of the ``power'' in an image at a given spatial scale, closure phases are particularly sensitive to asymmetry; e.g., a point-symmetric image has identically zero closure phases.  These are consistent with a constant closure phase of $0^\circ$, with typical uncertainties of $10^\circ$.

\section{Results}
\label{results}

Predicted images are shown in Fig.~\ref{shadow_maps}, corresponding to the synchrotron intensity at 230 GHz from DM. The upper panels correspond to a DM spike with $\gamma_{\mathrm{sp}} = 7/3$, whereas the maps in the lower panels are computed for the no-spike case, for which we consider a standard Navarro-Frenk-White (NFW) DM profile with power-law index $\gamma = 1$. We assume annihilation of 10 GeV DM particles into $b\bar{b}$,\footnote{We focus on the standard $b$ quark channel for simplicity. Injection of electrons through a different channel would primarily result in a rescaling of the intensity at the frequency of interest, slightly changing the best-fit cross sections we derive.} and a magnetic field of 10 G, for a static BH (left panels) and a maximally rotating BH (right panels). 

The photon ring, i.e.~the bright ring of radius $\sim 20\ \mu \mathrm{as}$ that surrounds the darker shadow of the BH, is clearly visible in the simulations for all DM models we consider, although in practice in the absence of a spike the signal is too weak to be detectable with the EHT, as discussed in the following. The presence of a photon ring introduces small-scale structure into the signal, readily observable with the EHT on long baselines. For a static BH the shadow is exactly circular. For all but the most rapidly rotating BHs it is also very nearly circular \cite{Johannsen2010,Johannsen2013}. For a maximally rotating Kerr BH viewed from the equatorial plane the photon ring is flattened in the direction aligned with the BH spin.

At scales above 25~$\mu$as the DM spike-induced emission produces a diffuse synchrotron halo whose intensity falls with radius as a power law with index $\approx 3.5$.  This is generic, occurring independently of the BH spin and is present even when gravitational lensing is ignored.  The extended nature of this component ensures that it is subdominant on Earth-sized baselines. In the absence of a spike, the profile is much flatter, and falls with radius as a power law with index $\approx 1$. Figure~\ref{shadow_maps} also illustrates the fact that the intensity is significantly enhanced in the presence of a DM spike with respect to the no-spike case. To better stress this enhancement, we show the maps for the spike case computed for very small annihilation cross sections of a few $10^{-31}\ \mathrm{cm^{3}\ s^{-1}}$---corresponding to the best fits to the EHT data, as discussed below---while in the absence of a spike we use the thermal s-wave cross section of $3 \times 10^{-26}\ \mathrm{cm^{3}\ s^{-1}}$.\footnote{For a light candidate (10 GeV), the spike becomes detectable when the cross section is greater than $10^{-31}\ \mathrm{cm^3\ s^{-1}}$, while for a TeV candidate, the spike becomes visible when the cross section exceeds $10^{-27} \ \mathrm{cm^3\ s^{-1}}$. However, in both cases, a NFW cusp would lead to a much smaller emission and would be essentially invisible unless the cross section is about $10^{-17}\ \mathrm{cm^3\ s^{-1}}$ if $m_{\mathrm{DM}} \sim 10\ \mathrm{GeV}$ and $10^{-13}\ \mathrm{cm^3\ s^{-1}}$ if $m_{\mathrm{DM}} \sim 1\ \mathrm{TeV}$, which is completely excluded by indirect detection limits (e.g.~Refs.~\cite{Ackermann2015dwarfs6years,Abdallah2016}).}

\begin{figure*}[t!]
\centering
\includegraphics[width=0.49\linewidth]{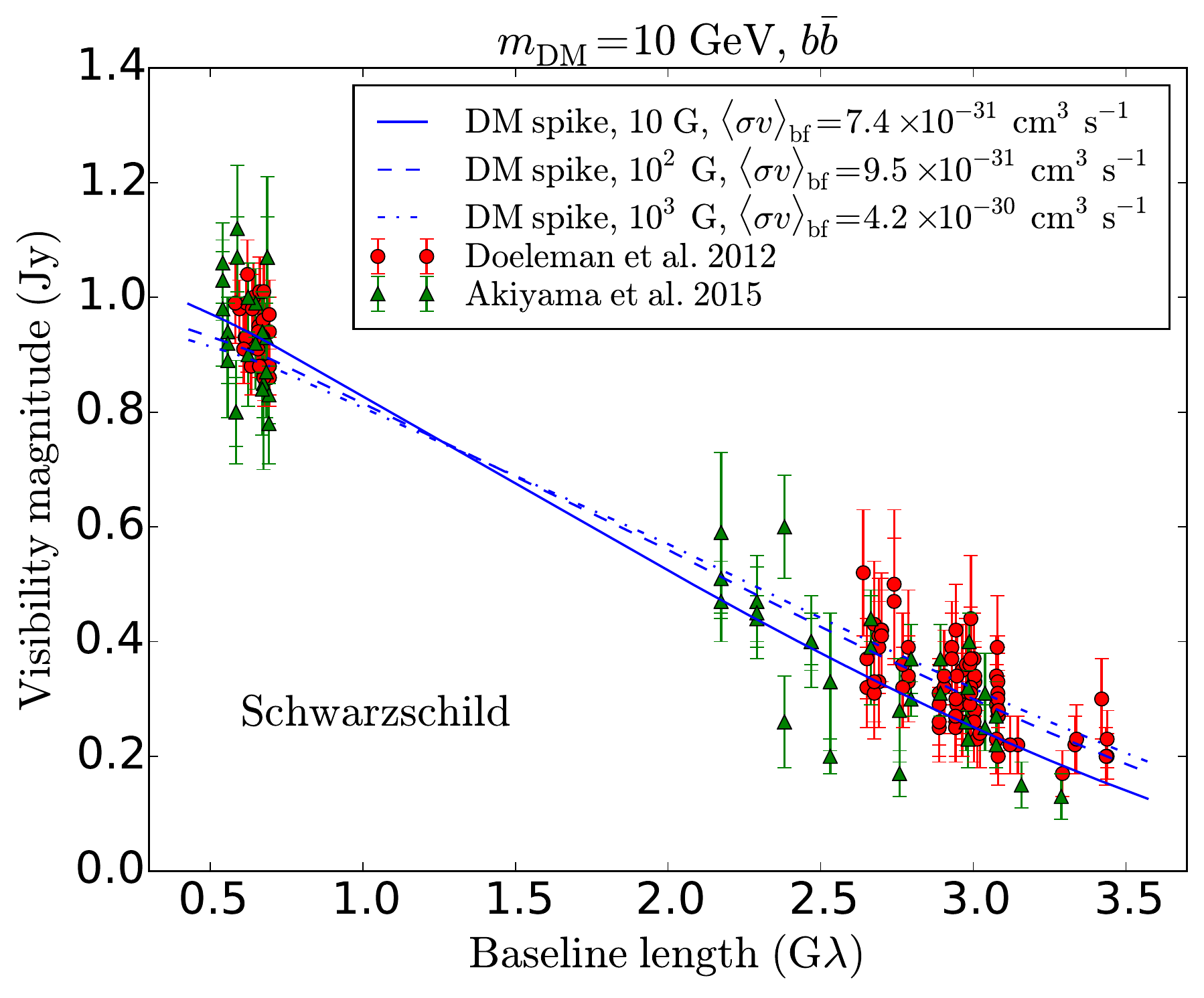} \includegraphics[width=0.49\linewidth]{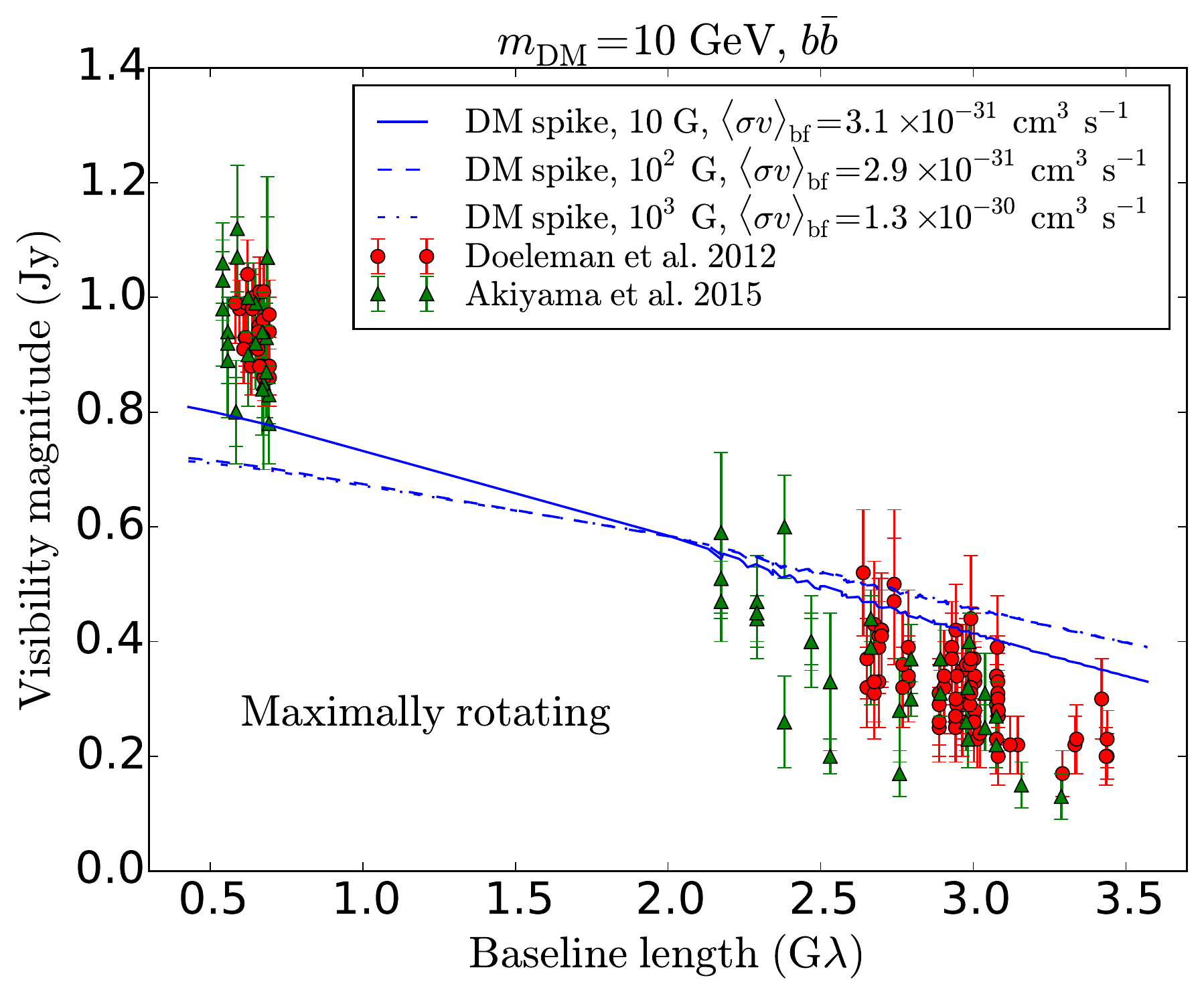}
\caption{\label{Visibility}Visibility amplitude at 230 GHz as a function of baseline length. The blue solid line represents the synchrotron emission from a spike of 10 GeV DM annihilating into $b\bar{b}$, with $B = 10\ \mathrm{G}$ (solid), $B = 10^{2}\ \mathrm{G}$ (dashed), and $B = 10^{3}\ \mathrm{G}$ (dot-dashed), for a Schwarzschild BH (left panel) and a maximally rotating BH (right panel). The annihilation cross sections correspond to the best fit to the EHT data from Refs.~\cite{Doeleman2012,Akiyama2015}, given in Tables \ref{table_best_fits_Schwarzschild} and \ref{table_best_fits_maximally_rotating} for the Schwarzschild and maximally rotating cases respectively.}
\end{figure*}

Shown in the blue solid line in Fig.~\ref{Visibility} is the visibility amplitude at 230 GHz as a function of baseline length for the current EHT triangle, for the simulated DM-induced synchrotron signal, computed with a cross section that gives the best fit to the EHT measurements from Refs.~\cite{Doeleman2012,Akiyama2015}. The left panel corresponds to the Schwarzschild case and the right panel to the maximally rotating case. The DM-induced visibility amplitudes shown in Fig.~\ref{Visibility} correspond to the intensity maps shown in Fig.~\ref{shadow_maps}. Note that no additional astrophysical component has been included in these. 

A standard NFW cusp actually results in visibility amplitudes that are about 8 orders of magnitude lower than the EHT data. Note that these no-spike visibility amplitudes are not shown in Fig.~\ref{Visibility} for clarity. Therefore, the EHT is only sensitive to spiky profiles, which makes it a dedicated probe of such sharply peaked DM distributions.\footnote{Moreover, the intensity profile for the no-spike case is much flatter than for a spiky profile, leading to a larger ratio of flux on long to short baselines, with no significant contribution on long baselines.}

\begin{table*}[t]
\begin{center}
\caption{\label{table_best_fits_Schwarzschild}Best-fit annihilation cross section and reduced chi-squared $\chi^{2}_{\mathrm{red}} = \chi^{2}/\mathrm{d.o.f.}$ for various DM masses and magnetic field strengths, for the Schwarzschild case. Values of the reduced chi-squared are given for illustration.}
\centering
\begin{tabular}{c|c|c|c}
\hline \hline
 & $m_{\mathrm{DM}} = 10\ \mathrm{GeV}$ & $m_{\mathrm{DM}} = 10^{2}\ \mathrm{GeV}$ & $m_{\mathrm{DM}} = 10^{3}\ \mathrm{GeV}$\\
 \hline
$B = 10\ \mathrm{G}$ & $\left\langle \sigma v \right\rangle_{\mathrm{bf}} = 7.4 \times 10^{-31}\ \mathrm{cm^{3}\ s^{-1}}$,  $\chi^{2}_{\mathrm{red}} = 1.4$ & $\left\langle \sigma v \right\rangle_{\mathrm{bf}} = 2.8 \times 10^{-29}\ \mathrm{cm^{3}\ s^{-1}}$,  $\chi^{2}_{\mathrm{red}} = 1.4$ & $\left\langle \sigma v \right\rangle_{\mathrm{bf}} = 1.2 \times 10^{-27}\ \mathrm{cm^{3}\ s^{-1}}$,  $\chi^{2}_{\mathrm{red}} = 1.4$\\
\hline
$B = 10^{2}\ \mathrm{G}$ & $\left\langle \sigma v \right\rangle_{\mathrm{bf}} = 9.5 \times 10^{-31}\ \mathrm{cm^{3}\ s^{-1}}$,  $\chi^{2}_{\mathrm{red}} = 1.5$ & $\left\langle \sigma v \right\rangle_{\mathrm{bf}} = 4.4 \times 10^{-29}\ \mathrm{cm^{3}\ s^{-1}}$,  $\chi^{2}_{\mathrm{red}} = 1.5$ & $\left\langle \sigma v \right\rangle_{\mathrm{bf}} = 1.8 \times 10^{-27}\ \mathrm{cm^{3}\ s^{-1}}$,  $\chi^{2}_{\mathrm{red}} = 1.5$\\
\hline
$B = 10^{3}\ \mathrm{G}$ & $\left\langle \sigma v \right\rangle_{\mathrm{bf}} = 4.2 \times 10^{-30}\ \mathrm{cm^{3}\ s^{-1}}$,  $\chi^{2}_{\mathrm{red}} = 1.8$ & $\left\langle \sigma v \right\rangle_{\mathrm{bf}} = 1.8 \times 10^{-28}\ \mathrm{cm^{3}\ s^{-1}}$,  $\chi^{2}_{\mathrm{red}} = 1.8$ & $\left\langle \sigma v \right\rangle_{\mathrm{bf}} = 8.1 \times 10^{-27}\ \mathrm{cm^{3}\ s^{-1}}$,  $\chi^{2}_{\mathrm{red}} = 1.7$ \\
\hline \hline
\end{tabular}
\end{center}
\end{table*}

\begin{table*}[t]
\caption{\label{table_best_fits_maximally_rotating}Best-fit annihilation cross section and reduced chi-squared $\chi^{2}_{\mathrm{red}} = \chi^{2}/\mathrm{d.o.f.}$ for various DM masses and magnetic field strengths, for the maximally rotating case. Values of the reduced chi-squared are given for illustration.}
\centering
\begin{tabular}{c|c|c|c}
\hline \hline
 & $m_{\mathrm{DM}} = 10\ \mathrm{GeV}$ & $m_{\mathrm{DM}} = 10^{2}\ \mathrm{GeV}$ & $m_{\mathrm{DM}} = 10^{3}\ \mathrm{GeV}$\\
 \hline
$B = 10\ \mathrm{G}$ & $\left\langle \sigma v \right\rangle_{\mathrm{bf}} = 3.1 \times 10^{-31}\ \mathrm{cm^{3}\ s^{-1}}$,  $\chi^{2}_{\mathrm{red}} = 6.5$ & $\left\langle \sigma v \right\rangle_{\mathrm{bf}} = 1.2 \times 10^{-29}\ \mathrm{cm^{3}\ s^{-1}}$,  $\chi^{2}_{\mathrm{red}} = 6.0$ & $\left\langle \sigma v \right\rangle_{\mathrm{bf}} = 5.2 \times 10^{-28}\ \mathrm{cm^{3}\ s^{-1}}$,  $\chi^{2}_{\mathrm{red}} = 5.8$\\
\hline
$B = 10^{2}\ \mathrm{G}$ & $\left\langle \sigma v \right\rangle_{\mathrm{bf}} = 2.9 \times 10^{-31}\ \mathrm{cm^{3}\ s^{-1}}$,  $\chi^{2}_{\mathrm{red}} = 11$ & $\left\langle \sigma v \right\rangle_{\mathrm{bf}} = 1.3 \times 10^{-29}\ \mathrm{cm^{3}\ s^{-1}}$,  $\chi^{2}_{\mathrm{red}} = 11$ & $\left\langle \sigma v \right\rangle_{\mathrm{bf}} = 5.6 \times 10^{-28}\ \mathrm{cm^{3}\ s^{-1}}$,  $\chi^{2}_{\mathrm{red}} = 11$\\
\hline
$B = 10^{3}\ \mathrm{G}$ & $\left\langle \sigma v \right\rangle_{\mathrm{bf}} = 1.3 \times 10^{-30}\ \mathrm{cm^{3}\ s^{-1}}$,  $\chi^{2}_{\mathrm{red}} = 12$ & $\left\langle \sigma v \right\rangle_{\mathrm{bf}} = 5.6 \times 10^{-29}\ \mathrm{cm^{3}\ s^{-1}}$,  $\chi^{2}_{\mathrm{red}} = 12$ & $\left\langle \sigma v \right\rangle_{\mathrm{bf}} = 2.5 \times 10^{-27}\ \mathrm{cm^{3}\ s^{-1}}$,  $\chi^{2}_{\mathrm{red}} = 12$ \\
\hline \hline
\end{tabular}
\end{table*}

As shown in Fig.~\ref{Visibility}, a spike of annihilating DM gives a good fit to the EHT measurements of the visibility amplitudes, with best-fit cross sections and reduced chi-squared---$\chi^{2}_{\mathrm{red}} = \chi^{2}/\mathrm{d.o.f.}$, with a number of degrees of freedom (d.o.f.) equal to 160 (data points) minus one (cross section)---given in Tables \ref{table_best_fits_Schwarzschild} and \ref{table_best_fits_maximally_rotating}.\footnote{The mass dependence of the intensity---and thus of the best-fit cross section---is fairly simple, though it changes from one channel to another. For the $b\bar{b}$ channel, the intensity goes roughly as $ \log(m_{\mathrm{DM}})/m_{\mathrm{DM}}^2$---the logarithm appears in the integral of the injection spectrum and $m_{\mathrm{DM}}^2$ in the number density of DM particles---which results in an increase of a factor $\sim 50$ in the cross section when increasing the mass by an order of magnitude. This allows for an extrapolation of our results at higher masses (typically 100 TeV), where prompt $\gamma$-ray emission is however more constraining \cite{M87_limits_my_paper}.} While the fits appear by eye to be quite good, the reduced chi-squareds coupled with the large number of degrees of freedom result in a p-value $< 0.002$, implying that some structural component is missing in our model.  This is not, in itself, surprising given the extraordinary simplicity of the DM spike model and our neglect of the contributions from the observed larger-scale radio emission associated with the jet.

The morphology of the predicted visibility amplitudes is only weakly sensitive to changes in the DM mass, the annihilation channel or the magnetic field, resulting primarily in different best-fit cross sections. A small increase in the ratio of visibilities on long baselines to those on short baselines for larger magnetic fields arises from the higher intrinsic synchrotron peak coupled with the gravitational redshift, which increases the 230~GHz emission near the horizon.

The visibility amplitudes are more sensitive to the characteristics of the BH. Rotating BHs exhibit larger relative visibility amplitudes on long baselines than static BHs as a consequence of the bright emission from comparatively smaller radii.  As a result, the quality of the fit is much better for a Schwarzschild BH.

\begin{figure}[h!]
\centering
\includegraphics[width=0.99\linewidth]{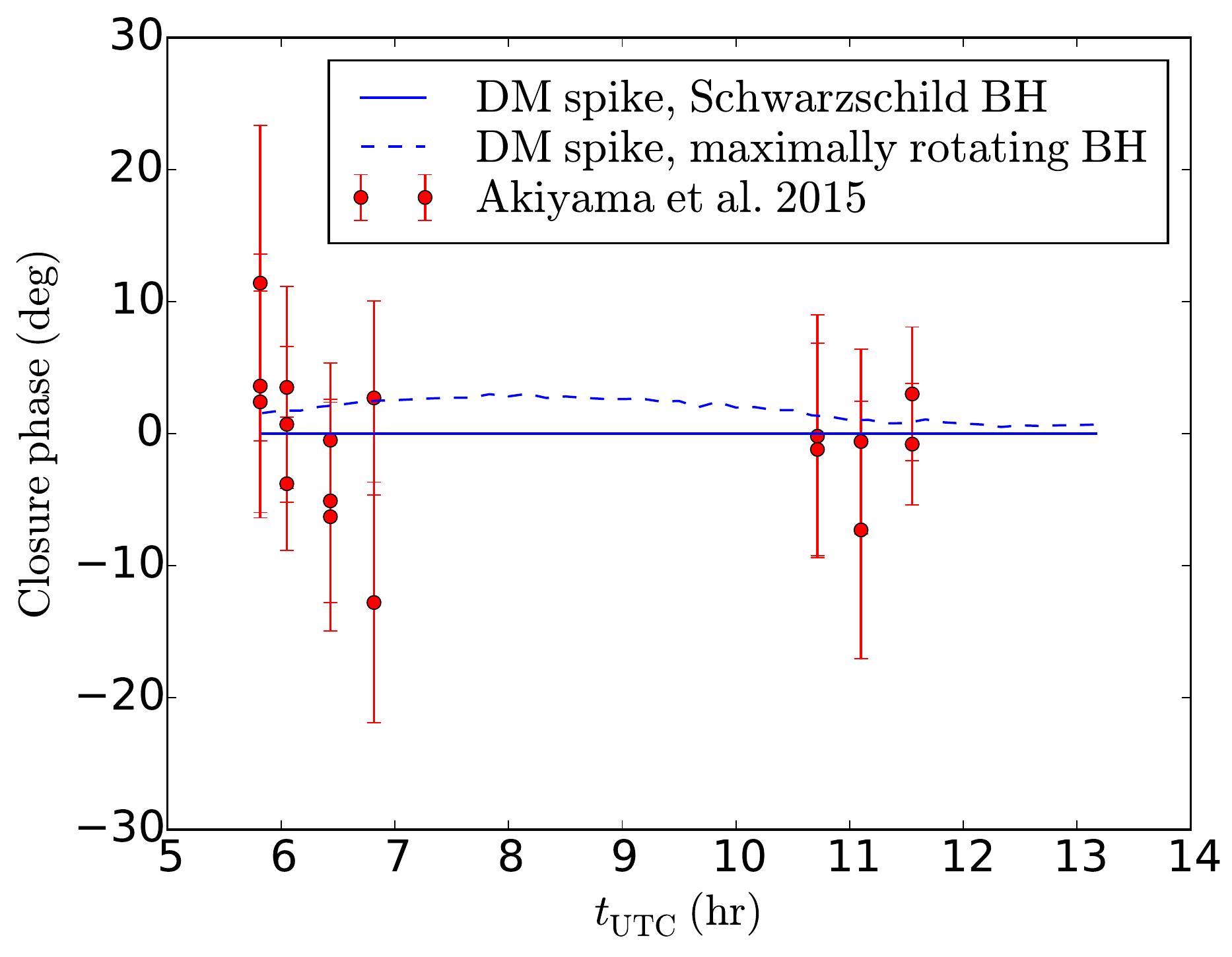} 
\caption{\label{closure_phases}Closure phase as a function of universal time, for a Schwarzschild BH (solid) and a maximally rotating BH (dashed), for the current VLBI triangle between Arizona, California and Hawaii.}
\end{figure}

The closure phases for the DM-spike-induced signal are shown in Fig.~\ref{closure_phases}, for the Schwarzschild case (solid line) and the maximally rotating case (dashed line). The symmetry of the simulated signal is insensitive to the various parameters so we do not need to specify them here.  As expected, the closure phase is identically zero for a Schwarzschild BH, while it is slightly larger than zero for the maximally rotating case. In both cases, closure phases for the DM-induced emission are consistent with the low closure phases observed.

Small closure phases are also typical of astrophysical models on the Hawaii-CARMA-SMT triangle \cite{Akiyama2015}.  This is because of the near degenerate nature of the projected baseline triangle due to the comparatively short CARMA-SMT baseline; closure phases on trivial triangles (in which one baseline has zero length) vanish identically.  However, the inclusion of a number of additional sites in the near future will result in many additional, open triangles for which the closure phases are likely to differ substantially from zero \cite{Akiyama2015}.  These will be instrumental to discriminating  both between astrophysical and DM-dominated models.

\section{Conclusion}
\label{conclusion}

In this paper, we have demonstrated the potential of the EHT for DM searches. We have presented the first model of the DM-induced synchrotron emission in the close vicinity of a SMBH accounting for strong gravitational lensing effects. Our conclusions are the following.
\begin{itemize}
\item The synchrotron emission from DM spikes should be readily visible in EHT images of M87 if present. This remains true even for very small values of the annihilation cross section.  The resulting emission follows the structure of the DM spike, resulting in a synchrotron halo extending from horizon scales to roughly $100\ \mu\mathrm{as}$.
\item Within the spike emission, the silhouette of the BH is clearly visible on small scales for all spins and models we considered.  This imparts small-scale structure on the image on scales of $50\ \mu\mathrm{as}$, and contributes substantially to the visibilities on long baselines.
\item DM spike emission provides an adequate fit to the existing horizon-scale structural constraints from the EHT.  This necessarily ignores astrophysical contributions associated with the jet launching region.  Such an additional astrophysical component is strongly motivated by the existence of extended emission at wavelengths of 3 mm and longer. Even within the 1.3 mm data, the fit quality suggests that the simple spike structures we present here are incomplete.
\item Nevertheless, the limits on M87's flux and small-scale structure place corresponding constraints on the putative DM annihilation cross sections.  For a 10 GeV DM candidate this cross section must be less than a few $10^{-31}\ \mathrm{cm^3\ s^{-1}}$, close to the characteristic cross sections for p-wave-suppressed annihilation. The introduction of additional astrophysical components would decrease this limit further.
\item EHT observations in the near future will include a number of additional stations, enabling the reconstruction of M87's image with substantially higher fidelity.  As a result, the limits on the existence of DM spikes and the properties of DM candidates will similarly improve. Thus, the EHT opens a new, powerful path to probing the structure and features of DM in the centers of galaxies.
\end{itemize}

\acknowledgments{This work has been supported by UPMC, CNRS and STFC. This research has also received support at IAP from the ERC Project No.~267117 (DARK) hosted by UPMC, and has been carried out in the ILP LABEX (ANR-10-LABX-63) and supported by French state funds managed by the ANR, within the Investissements d'Avenir programme (ANR-11-IDEX-0004-02). A.E.B. and M.K.~receive financial support from the Perimeter Institute for Theoretical Physics and the Natural Sciences and Engineering Research Council of Canada through a Discovery Grant.  Research at Perimeter Institute is supported by the Government of Canada through Industry Canada and by the Province of Ontario through the Ministry of Research and Innovation. We acknowledge support by the Canadian Institute for Advanced Research for the participation of A.E.B. and J.S. at the annual meetings of the CIFAR cosmology program where this project was initiated.}

\appendix

\section{Normalization of the DM profile}
\label{spike_normalization_M87}

Here we describe how we normalize the profile corresponding to a DM spike growing adiabatically from an initial power-law profile $\rho_{0} \left( r/r_0 \right)^{-\gamma}$ \cite{spikeGS} at the center of the M87 galaxy:
\begin{equation}
\rho (r) =
\begin{cases}
0 & r < R_{\mathrm{S}} \\
\dfrac{\rho_{\mathrm{sp}}(r) \rho_{\mathrm{sat}}}{\rho_{\mathrm{sp}}(r) + \rho_{\mathrm{sat}}} & R_{\mathrm{S}} \leq r < R_{\mathrm{sp}} \\
\rho_{0} \left( \dfrac{r}{r_0} \right)^{-\gamma} \left( 1 + \dfrac{r}{r_0} \right) ^{-2} & r \geq R_{\mathrm{sp}},
\end{cases}
\end{equation}
where the saturation density determined by DM annihilations reads
\begin{equation}
\label{saturation_density_appendix}
\rho_{\mathrm{sat}} = \dfrac{m_{\mathrm{DM}}}{\left\langle \sigma v \right\rangle t_{\mathrm{BH}}},
\end{equation}
where $m_{\mathrm{DM}}$ and $\left\langle \sigma v \right\rangle$ are respectively the mass and annihilation cross section of the DM particle, and we take $t_{\mathrm{BH}} = 10^{8}\ \mathrm{yr}$ for the age of the BH.\footnote{For an isotropic DM distribution function, a weak cusp going as $r^{-1/2}$ arises instead of a plateau \cite{Shapiro2016}.} The spike profile reads
\begin{equation}
\rho_{\mathrm{sp}}(r) = \rho_{\mathrm{R}} \left( \dfrac{R_{\mathrm{sp}}}{r} \right)^{\gamma_{\mathrm{sp}}},
\end{equation}
where $\rho_{\mathrm{R}} = \rho_{0} \left( R_{\mathrm{sp}}/r_0 \right) ^{-\gamma}$, $R_{\mathrm{sp}} = \alpha_{\gamma} r_0 \left( M_{\mathrm{BH}}/(\rho_{0}r_0^{3}) \right) ^{\frac{1}{3-\gamma}}$ and $\gamma_{\mathrm{sp}} = (9-2\gamma)/(4-\gamma)$. We use $M_{\mathrm{BH}} = 6.4 \times 10^9 M_{\odot}$ for the mass of the BH \citep{Gebhardt2009}, the corresponding Schwarzschild radius is $R_{\mathrm{S}} = 6 \times 10^{-4}\ \rm pc$, and we take $\alpha_{\gamma} = 0.1$. We fix $r_0 = 20\ \rm kpc$ for the halo (as for the Milky Way), and we must then determine the normalization $\rho_{0}$. 

We choose $\rho_{0}$ in such a way that the profile is compatible with both the total mass of the galaxy and the mass enclosed within the radius of influence of the BH, of order $10^5 R_{\mathrm{S}}$. We thus follow the procedure described in Ref.~\cite{Gorchtein_DM_AGN_jet}: the DM mass within the region that is relevant for the determination of the BH mass, typically within $R_0 = 10^5 R_{\mathrm{S}}$, must be smaller than the uncertainty on the BH mass $\Delta M_{\mathrm{BH}}$. $\rho_0$ is thus obtained by solving the following equation: 
\begin{equation}
\label{normalization}
\int_{R_{\mathrm{S}}}^{10^5 R_{\mathrm{S}}} \! 4 \pi r^{2} \rho(r) \,\mathrm{d}r = \Delta M_{\mathrm{BH}},
\end{equation}
with $\Delta M_{\mathrm{BH}} = 5 \times 10^8\ M_\odot$. Considering the complex dependence of $\rho$ on $\rho_0$, we use the fact that the mass is dominated by the contribution from $r \gg R_{\mathrm{S}}$, i.e., typically $r > R_{\mathrm{min}} = \mathcal{O}(100 R_{\mathrm{S}})$. In this regime we have $\rho \sim \rho_{\mathrm{sp}}(r)$. We can also factorize the dependence on $\rho_0$ in $\rho_{\mathrm{sp}}$, $\rho_{\mathrm{sp}}(r) = g_{\gamma}(r) \rho_0^{\frac{1}{4-\gamma}} \left( R_{\mathrm{sp}}'/r_0 \right) ^{-\gamma} \left( R_{\mathrm{sp}}'/r \right) ^{\gamma_{\mathrm{sp}}}$, with $R_{\mathrm{sp}}' = \alpha_{\gamma} r_{0} \left( M_{\mathrm{BH}}/r_{0}^{3}\right)^{\frac{1}{3-\gamma}}$, and we finally obtain
\begin{equation}
\rho_0 = \left( \dfrac{\left( 3-\gamma_{\mathrm{sp}}\right) \Delta M_{\mathrm{BH}}}{4 \pi R_{\mathrm{sp}}'^{\gamma_{\mathrm{sp}} - \gamma} r_0^{\gamma} \left( R_0^{3-\gamma_{\mathrm{sp}}} - R_{\mathrm{min}}^{3-\gamma_{\mathrm{sp}}} \right) } \right) ^{4-\gamma}.
\end{equation}
We take $\gamma = 1$, which corresponds to the NFW profile. The corresponding spike has a power-law index of $\gamma_{\mathrm{sp}} = 7/3$. Numerically, we get $\rho_0 \approx 2.5\ \rm GeV\ cm^{-3}$ for $\gamma = 1$. Finally, the total mass within 50 kpc is $\sim 4 \times 10^{12}\ M_{\odot}$, compatible with the value derived from observations, $6 \times 10^{12}\ M_{\odot}$ \citep{Merritt1993}.

In practice, the saturation radius $r_{\mathrm{sat}}$---for which $\rho_{\mathrm{sat}} = \rho(r_{\mathrm{sat}})$---is smaller than the Schwarzschild radius of the BH for all values of the DM mass and annihilation cross section of interest here, so that the DM profile reads more simply
\begin{equation}
\rho (r) =
\begin{cases}
0 & r < R_{\mathrm{S}} \\
\rho_{\mathrm{sp}}(r) & R_{\mathrm{S}} \leq r < R_{\mathrm{sp}} \\
\rho_{0} \left( \dfrac{r}{r_0} \right)^{-\gamma} \left( 1 + \dfrac{r}{r_0} \right) ^{-2} & r \geq R_{\mathrm{sp}}.
\end{cases}
\end{equation}
Note that one usually assumes that the DM profile vanishes below $4 R_{\mathrm{S}}$ (or $2 R_{\mathrm{S}}$ from the full relativistic calculation for a static BH \cite{Sadeghian2013}) due to DM particles captured by the BH. Here, for simplicity, to study the potential of the EHT for probing very steep power-law density profiles, we consider a DM spike that goes all the way down to the horizon of the BH, i.e.~$R_{\mathrm{S}}$ for a Schwarzschild BH and $R_{\mathrm{S}}/2$ for a maximally rotating BH. However, this simplification has a negligible impact on our results.

\section{Solving the cosmic-ray equation in the presence of an advection flow towards the central BH}
\label{advection_resolution}

The propagation equation of electrons and positrons in the presence of advection and synchrotron losses,
\begin{equation}
v\dfrac{\partial f}{\partial r} - \dfrac{1}{3 r^{2}} \dfrac{\partial}{\partial r}\left( r^{2} v \right) p \dfrac{\partial f}{\partial p} + \dfrac{1}{p^{2}} \dfrac{\partial}{\partial p} \left( p^{2} \dot{p} f \right) = Q,
\end{equation}
can be rewritten as
\begin{equation}
\dfrac{\partial f}{\partial r} + \dfrac{\dot{p}_{\mathrm{ad}} + \dot{p}_{\mathrm{syn}}}{v} \dfrac{\partial f}{\partial p} = - \dfrac{1}{v p^{2}} \dfrac{\partial}{\partial p} \left( p^{2} \dot{p}_{\mathrm{syn}} \right) f + \dfrac{Q}{v},
\end{equation}
where $\dot{p}_{\mathrm{ad}}$ is the momentum gain rate due to adiabatic compression in the advection process, and $v(r)$ is the velocity field of the accretion flow. The associated characteristic curves are obtained by solving the following differential equation:
\begin{equation}
\label{characteristic_ode}
\dfrac{\mathrm{d} p}{\mathrm{d} r} = \dfrac{\dot{p}_{\mathrm{ad}} + \dot{p}_{\mathrm{syn}}}{v}.
\end{equation}
Generalizing the method of Ref.~\cite{Aloisio2004} to an arbitrary power-law profile for the magnetic field, $B(r) = B_{0} \left( r/R_{\mathrm{S}} \right)^{-\alpha/2}$, solving Eq.~\eqref{characteristic_ode} with the initial condition $p(R_{\mathrm{inj}}) = p_{\mathrm{inj}}$ leads to 
\begin{align}
\label{characteristic}
& p(r;R_{\mathrm{inj}},p_{\mathrm{inj}}) = p_{\mathrm{inj}} \nonumber \\
& \times \left[ \dfrac{k_{0} R_{\mathrm{S}}^{\alpha-\frac{1}{2}}}{(\alpha-1) c} r^{\frac{3}{2}-\alpha} p_{\mathrm{inj}} \left( 1 - \left( \dfrac{r}{R_{\mathrm{inj}}}\right)^{\alpha-1} \right) + \left(\dfrac{r}{R_{\mathrm{inj}}}\right)^{\frac{1}{2}} \right]^{-1},
\end{align}
where 
\begin{equation}
k_{0} = \dfrac{2 \sigma_{\mathrm{T}} B_{0}^{2}}{3 \mu_{0} (m_{\mathrm{e}} c)^{2}}.
\end{equation}
We consider the case $\alpha = 0$, corresponding to a homogeneous magnetic field. 

The solution of the propagation equation in the ultrarelativistic regime is then given by
\begin{equation}
\label{solution}
f(r,p) = \dfrac{1}{c} \left( \dfrac{r}{R_{\mathrm{S}}} \right) \int_{r}^{r_{\mathrm{acc}}} \! Q(R_{\mathrm{inj}},p_{\mathrm{inj}}) \left( \dfrac{R_{\mathrm{inj}}}{R_{\mathrm{S}}} \right)^{\frac{5}{2}} \left( \dfrac{p_{\mathrm{inj}}}{p} \right)^{4} \, \mathrm{d}R_{\mathrm{inj}},
\end{equation}
where $p_{\mathrm{inj}} \equiv p_{\mathrm{inj}}(R_{\mathrm{inj}};r,p)$ is the injection momentum of an electron injected at $R_{\mathrm{inj}}$ ($\geq r$) and arriving at $r$ with momentum $p$. Using Eq.~\eqref{characteristic} and expressing $p_{\mathrm{inj}}$ as a function of $p$, we obtain, for $\alpha = 0$
\begin{equation}
\label{p_inj}
p_{\mathrm{inj}}(R_{\mathrm{inj}};r,p) = p \left[ \dfrac{k_{0} R_{\mathrm{S}}^{-\frac{1}{2}}}{c} R_{\mathrm{inj}}^{\frac{3}{2}} p \left( \dfrac{r}{R_{\mathrm{inj}}} -1 \right) + \left( \dfrac{R_{\mathrm{inj}}}{r} \right)^{\frac{1}{2}} \right] ^{-1}.
\end{equation}
Note that the denominator of $p_{\mathrm{inj}}$ can vanish and become negative, leading to nonphysical values of the injection momentum. This is related to the efficiency of the accretion flow and characterizes the region of the injection parameters $(R_{\mathrm{inj}},p_{\mathrm{inj}})$ corresponding to a given arrival point $(r,p)$. In practice, $p_{\mathrm{inj}}$ remains positive for $R_{\mathrm{inj}} < R_{\mathrm{inj}}^{0}$ where
\begin{equation}
R_{\mathrm{inj}}^{0} = r + \dfrac{c}{k_{0} p} \left( \dfrac{r}{R_{\mathrm{S}}} \right)^{-\frac{1}{2}}.
\end{equation}
We then use this value as an effective upper bound for the integral of Eq.~\eqref{solution}.

\bibliographystyle{apsrev4-1} 
\bibliography{/Users/thomaslacroix/Documents/bibliography/biblio_bibtex/biblio}

\end{document}